\documentclass[twocolumn,showpacs,aps,prb,superscriptaddress,floatfix]{revtex4}
\usepackage{graphicx}
\usepackage{dcolumn}
\usepackage{bm}
\begin{document}
\preprint{}
%

\title{Singlet-Triplet Excitations in the Unconventional Spin-Peierls System TiOBr}

\author{J.P. Clancy}
\affiliation{Department of Physics and Astronomy, McMaster University,
Hamilton, Ontario, L8S 4M1, Canada}

\author{B.D. Gaulin}
\affiliation{Department of Physics and Astronomy, McMaster University,
Hamilton, Ontario, L8S 4M1, Canada}
\affiliation{Brockhouse Institute for Materials Research, McMaster University,
Hamilton, Ontario, L8S 4M1, Canada}
\affiliation{Canadian Institute for Advanced Research, 180 Dundas St. W.,
Toronto, Ontario, M5G 1Z8, Canada}

\author{C.P. Adams}
\affiliation{Department of Physics, St. Francis Xavier University,
Antigonish, Nova Scotia, B2G 2W5, Canada}

\author{G.E. Granroth}
\affiliation{Neutron Scattering Sciences Division, Oak Ridge National Laboratory, Oak Ridge, Tennessee 37831, USA}

\author{A.I. Kolesnikov}
\affiliation{Neutron Scattering Sciences Division, Oak Ridge National Laboratory, Oak Ridge, Tennessee 37831, USA}

\author{T.E. Sherline}
\affiliation{Neutron Scattering Sciences Division, Oak Ridge National Laboratory, Oak Ridge, Tennessee 37831, USA}

\author{F.C. Chou}
\affiliation{Center for Condensed Matter Sciences, National Taiwan University, Taipei 106, Taiwan.}

\begin{abstract}
We have performed time-of-flight neutron scattering measurements on powder samples of the unconventional spin-Peierls compound TiOBr using the fine-resolution Fermi chopper spectrometer (SEQUOIA) at the SNS.  These measurements reveal two branches of magnetic excitations within the commensurate and incommensurate spin-Peierls phases, which we associate with n = 1 and n = 2 triplet excitations out of the singlet ground state.  These measurements represent the first direct measure of the singlet-triplet energy gap in TiOBr, which is determined to be $E_{g}$ = 21.2$\pm$1.0 meV.   

\end{abstract}

\maketitle

TiOBr belongs to a select family of quasi-one-dimensional (1D) magnetic materials which undergo a spin-Peierls phase transition.  These systems are characterized by a combination of spin 1/2 magnetic moments, short-range antiferromagnetic interactions, and strong magnetoelastic coupling\cite{Spin-Peierls}.  At the spin-Peierls transition temperature, T$_{SP}$, the quasi-1D spin chains in these systems distort and dimerize, leading to the formation of a non-magnetic singlet ground state.  Very few materials have been found to exhibit a spin-Peierls phase transition, and to date there are only three known inorganic spin-Peierls systems: CuGeO$_{3}$, TiOCl, and TiOBr.

The isostructural Ti$^{3+}$-based spin-Peierls compounds, TiOCl and TiOBr, have recently attracted considerable attention as they have been shown to deviate from the standard spin-Peierls picture in several important respects.  TiOCl and TiOBr exhibit not one, but two successive phase transitions upon cooling - a continuous transition from a uniform paramagnetic phase to an incommensurate spin-Peierls state at T$_{C2}$ $\sim$ 92K/48K, followed by a discontinuous transition into a commensurate spin-Peierls state at T$_{C1}$ $\sim$ 65K/27K\cite{Seidel, Kato, Lemmens_TiOBr, Smaalen, Sasaki_JPSJ, Imai, Baker, Lemmens_TiOCl, Caimi, Sasaki_unpublished, Ruckamp, Clancy_TiOCl, Clancy_TiOBr}.  These compounds are also distinguished from conventional spin-Peierls systems by their unusually high transition temperatures and the surprisingly large energy gap ($E_{g}$) between the singlet ground state and the first triplet excited state\cite{Imai, Baker, Lemmens_TiOCl, Caimi, Sasaki_unpublished}.  In TiOCl the spin-Peierls state has also been shown to be particularly sensitive to the presence of quenched non-magnetic impurities\cite{Clancy_TiScOCl}.

Experimental studies of these systems have typically focused on TiOCl rather than TiOBr.  In practice, this has been the case because TiOBr is extremely hygroscopic (even more so than TiOCl) and is very difficult to synthesize in high quality single crystal form.  NMR\cite{Imai}, $\mu$SR\cite{Baker}, Raman\cite{Lemmens_TiOCl} and IR spectroscopy\cite{Caimi} measurements on TiOCl present a consistent picture of a singlet-triplet energy gap which is between 430 - 440 K.  Similar measurements have not been performed for TiOBr, although a gap of $\sim$ 149 K = 12.6 meV has been inferred from low temperature magnetic susceptibility measurements\cite{Sasaki_unpublished}.  To date, neutron scattering measurements have not been reported for either TiOCl or TiOBr due to the limited size of available single crystal samples.  As a result, there is a surprising lack of information regarding the magnetic excitation spectrum of these unconventional spin-Peierls systems.
 
In this letter we report inelastic neutron scattering measurements on a powder sample of TiOBr.  These measurements reveal the magnetic excitation spectrum of the system, providing the first direct measure of the singlet-triplet energy gap in this unconventional spin-Peierls material.  We observe two sets of magnetic excitations, at $\Delta$E $\sim$ 21 meV and $\Delta$E $\sim$ 41 meV, which we associate with n = 1 and n = 2 triplet excitations, respectively.  Our measurements show that the bandwidth of these excitations is relatively narrow compared to the size of the singlet-triplet energy gap, suggesting that the excitations are fairly well-localized in nature.  Furthermore, from the energy scales of the n = 1 and n = 2 triplet excitations we can infer that interactions between excited triplets are small.

Time-of-flight neutron scattering measurements were performed on a 2.85 g powder sample of TiOBr.  Due to the volatile nature of TiBr$_{4}$, the sample was prepared by mixing TiO$_{2}$:Ti:TiBr$_{4}$ powder in a 2:1:1.4 molar ratio and packing the material in quartz tubing inside a glove box.  The packed quartz tubing was removed from the glove box using a valve-controlled transfer tube and was gradually evacuated before flame sealing.  The sealed quartz tube was then heated at 650 - 700 C for 20 hours, resulting in a final product which was mainly polycrystalline in form, with minor visible crystal formation.

Neutron scattering measurements were performed using SEQUOIA, the recently commissioned fine-resolution Fermi chopper spectrometer at the Spallation Neutron Source (SNS) at Oak Ridge National Laboratory (ORNL)\cite{SEQUOIA_PhysicaB, SEQUOIA_JPCS}.  SEQUOIA is an ideal instrument for the study of weak magnetic scattering, as it offers a combination of high neutron flux and excellent low-{\bf Q} detector coverage.  Measurements were carried out with Fermi chopper 2 phased for an incident energy of E$_{i}$ = 60 meV and rotating at a frequency of 480 Hz.  This chopper provides $\sim$ 1.5 meV energy resolution at the elastic line for these conditions.  A T0 chopper, used to eliminate unwanted high energy neutrons, was operated at 90 Hz.

The magnetic excitation spectrum of TiOBr is illustrated by the maps of inelastic neutron scattering intensity, S(Q,E), provided in Fig. 1.  The magnetic scattering in TiOBr is expected to be quite weak, due to both the small size (S = 1/2) and low density (one per formula unit) of the magnetic moments in the system.  This problem is exacerbated by the powder nature of the sample, which causes the weak magnetic signal to be averaged over all possible directions in reciprocal space.  A further complication arises from the fact that the magnetic excitations in TiOBr occur close in energy to much stronger phonon modes associated with TiOBr and the aluminum sample environment.  This combination of factors makes it challenging to isolate the inelastic magnetic scattering in TiOBr, and makes it critical that the experimental background is carefully analyzed and understood.  

\begin{figure}
\includegraphics{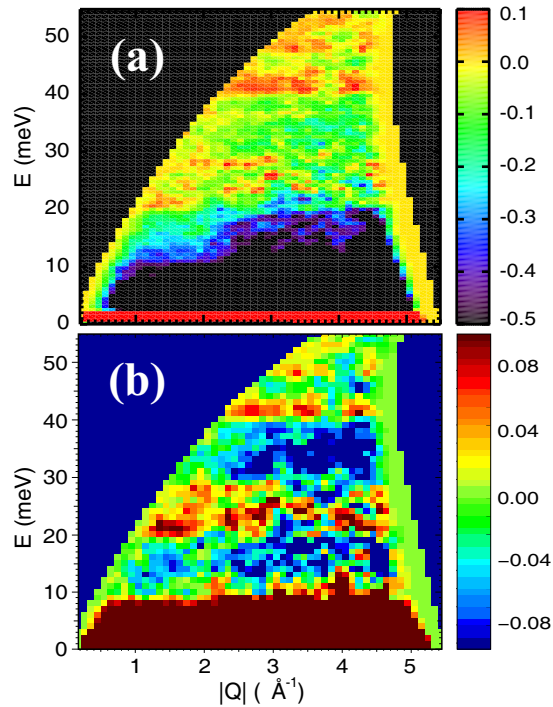}
\caption{(Color online) Color contour maps of the inelastic neutron scattering intensity, S(Q,E), observed in TiOBr.  The low temperature magnetic scattering is illustrated by: (a) A difference map comparing the scattering at T = 8 K (within the commensurate spin-Peierls state) to the scattering at T = 80K (within the paramagnetic state).  (b) A difference map of the T = 8 K and T = 80 K data sets where the high temperature subtraction has been weighted by an appropriate Bose correction, as described in the text.}
\end{figure}

In general, the scattering observed at any given temperature will consist of three terms: magnetic scattering, phonon scattering, and an approximately temperature independent background resulting from the sample environment, detector dark current, etc.  The background term can effectively be eliminated by performing an empty can background subtraction, leaving only the magnetic and phonon contributions from the sample.  The magnetic scattering can then be isolated from the phonon scattering by taking advantage of the different temperature dependencies of the two terms and performing a high temperature background subtraction.  At high temperatures, for T $>$ T$_{C2}$, the phonon scattering should dominate the magnetic scattering, and we can make the approximation that:
\begin{eqnarray}
\nonumber
I(\omega,T_{high})&=&I_{ph}(\omega,T_{high})+I_{mag}(\omega,T_{high}) \\
&\approx&I_{ph}(\omega,T_{high})
\end{eqnarray}
The low temperature magnetic scattering can then be determined from:
\begin{eqnarray}
\nonumber
I_{mag}(\omega,T_{low}) = I(\omega,T_{low}) - I_{ph}(\omega,T_{low}) \\
= I(\omega,T_{low}) - \left[\frac{1-e^{\left(-\hbar\omega/k_{b}T_{high}\right)}}{1-e^{\left(-\hbar\omega/k_{b}T_{low}\right)}} \right] I(\omega,T_{high})
\end{eqnarray}

The color contour maps provided in Fig. 1 show the inelastic magnetic scattering at base temperature (T = 8 K).  Fig. 1(a) is a simple difference map comparing the scattering at T = 8 K (within the commensurate spin-Peierls phase) and T = 80 K (within the uniform paramagnetic phase), while Fig. 1(b) is a similar map employing a Bose-weighted high temperature subtraction as described above.  Note that two bands of positive scattering intensity can be observed at energy transfers of $\sim$ 21 meV and $\sim$ 41 meV.  These bands represent magnetic excitations, which we associate with the n = 1 and n = 2 triplet excited states, respectively.  In addition, there are two regions of negative scattering intensity, from 8 meV to 20 meV and from 28 meV to 40 meV, which we associate with well-defined singlet-triplet energy gaps in the excitation spectrum.  

Fig. 2 shows a series of cuts through S(Q,E), where scattering intensity has been integrated over two different regions of {\bf Q} in order to examine the energy dependence of the inelastic scattering.  The {\bf Q}-windows have been chosen to highlight the bottom of the n = 1 triplet excitation (by integrating from {\bf Q} = 1 to 2 {\AA}$^{-1}$) and the n = 2 triplet excitation (by integrating from {\bf Q} = 2 to 3 {\AA}$^{-1}$).  The first set of cuts, shown in panels (a) and (b), were taken through S(Q,E) after performing an empty can background subtraction.  This eliminates scattering from the Al sample can, and in particular the strong Al phonon mode at $\sim$ 18 meV.  The cuts in panels (c) and (d) were taken through S(Q,E) after performing a high temperature (T = 80 K) background subtraction, while those in (e) and (f) were taken through S(Q,E) after a high temperature subtraction weighted by appropriate Bose factors.  As a result, the scattering intensity in panels (c) to (f) is almost entirely magnetic in origin. The solid lines provided in panels (c) to (f) represent fits to the data using multiple Lorentzian lineshapes.

\begin{figure}
\includegraphics{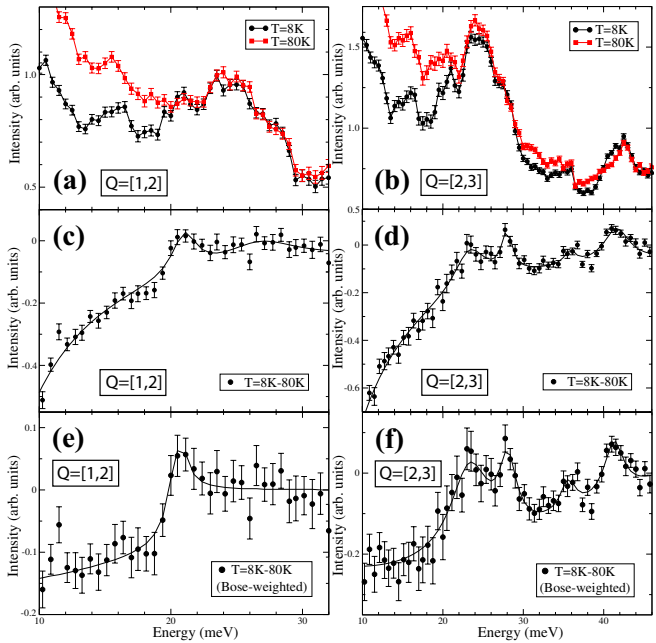}
\caption{(Color online) Representative energy cuts, I(E), taken through S(Q,E) at intervals of $|${\bf Q}$|$ = 1 to 2 {\AA}$^{-1}$ (left) and 2 to 3 {\AA}$^{-1}$ (right).  Panels (a) and (b) provide a direct comparison of the scattering at T = 8 K (within the commensurate spin-Peierls phase) and T = 80 K (within the paramagnetic phase).  An empty can background has been subtracted from (a) and (b) to eliminate scattering from the sample environment.  Panels (c) and (d) show cuts through the T = 8 K data set after a high temperature (T = 80 K) background subtraction has been used to isolate the magnetic scattering.  Panels (e) and (f) show similar cuts through the T = 8 K data set after a Bose-corrected high temperature background subtraction has been performed.}
\end{figure}

The magnitude of the singlet-triplet energy gap in TiOBr is defined by the lower bound of the n = 1 triplet excitation.  Thus, by fitting the data in Fig. 2 we obtain a value of $E_{g}$ = 21.2 $\pm$ 1.0 meV $\sim$ 250 K.  This value is significantly higher than the reported value of 12.6 meV inferred from magnetic susceptibility measurements\cite{Sasaki_unpublished}.  However, our experimental value of $E_{g}$ is remarkably consistent with previous measurements of the singlet-triplet energy gap in TiOCl\cite{Imai, Baker, Lemmens_TiOCl, Caimi}.  By starting from the reported value of $E_{TiOCl}$ = 430 - 440 K, and scaling by the ratio of the exchange couplings determined from magnetic susceptibility ($J_{TiOCl}$ = 660 - 676 K\cite{Seidel, Ruckamp} and $J_{TiOBr}$ = 364 - 376 K\cite{Kato, Ruckamp, Sasaki_JPSJ, Sasaki_unpublished}) one obtains a prediction of $E_{TiOBr}$ = $\left(\frac{J_{TiOBr}}{J_{TiOCl}}\right) \times E_{TiOCl}$ = 19.6 - 21.3 meV, in excellent agreement with our experimental results.  

As in the case of TiOCl, this value of $E_{g}$ is unusually large compared to both the size of the gap in other spin-Peierls systems ($E_{CuGeO_{3}}$ = 2.1 meV\cite{Nishi_CGO, Regnault_CGO}) and the size of the energy scale determined by T$_{C1}$ and T$_{C2}$.  The BCS prediction for $E_{g}$ in a conventional spin-Peierls system yields a value of $\frac{2E_{g}}{k_{B}T_{SP}}$ $\sim$ 3.5.  While this prediction is almost perfectly realized in the case of CuGeO$_{3}$ ($\frac{2E_{g}}{k_{B}T_{SP}}$ = 3.54\cite{Nishi_CGO, Regnault_CGO}), it appears to provide a poor description of the energy gap in the Ti-based spin-Peierls compounds ($\frac{2E_{g}}{k_{B}T_{SP}}$ $\sim$ 10 to 13 in TiOCl and 10 to 18 in TiOBr).

It is interesting to note that the energy scale for the n = 2 triplet excitation (41 meV) is almost exactly twice the energy scale of the n = 1 triplet excitation (21 meV).  This implies that the interactions between excited (S=1) triplets must be small, as any inter-triplet coupling should act to shift the n = 2 excitation away from $\Delta$E = 2$E_{g}$.  A similar result has been observed in CuGeO$_{3}$, where inelastic neutron scattering measurements reveal well-defined n = 1 triplet excitations at $\sim$ 2.1 meV which are separated from a continuum of states by a second, approximately equal, energy gap of $\sim$ 2 meV\cite{Ain_CGO, Arai_CGO}.  This may be contrasted with other singlet ground state systems, such as the Shastry-Sutherland system SrCu$_{2}$(BO$_{3}$)$_{2}$, in which the interactions between triplets are much stronger.  In SrCu$_{2}$(BO$_{3}$)$_{2}$, these interactions reduce the energy of the n = 2 triplet excitation by $\sim$ 40 \%, giving rise to n = 1 and n = 2 excitations at energies of $\sim$ 3 meV and 4.9 meV, respectively\cite{Gaulin_SCBO, Kageyama_SCBO}.

It is also instructive to consider the bandwidth of the magnetic excitations in TiOBr.  While it is difficult to determine a precise bandwidth due to the strength of the magnetic signal and the effects of powder averaging, it is reasonable to place an upper bound of $\sim$ 8 meV on the bandwidth of the n = 1 triplet excitation, which extends at most from 20 meV to 28 meV.  This bandwidth is relatively small compared to the size of the energy gap ($\sim$ 40 \% of $E_{g}$), suggesting that the triplet excitations in TiOBr are fairly well-localized in nature.  This represents a significant difference from the excitation spectrum of CuGeO$_{3}$, in which the dispersion ranges from $\sim$ 25 \% (inter-chain) to $\sim$ 800 \% (intra-chain) of the singlet-triplet energy gap\cite{Regnault_CGO, Arai_CGO}.  In part, the magnetic excitations in TiOBr may be less dispersive because of the geometric frustration inherent to the buckled Ti-O bilayers of the crystal structure.   Certainly geometric frustration is believed to be responsible for the largely dispersionless singlet-triplet excitations observed in SrCu$_{2}$(BO$_{3}$)$_{2}$\cite{Gaulin_SCBO, Kageyama_SCBO}. 

The temperature dependence of the magnetic excitation spectrum in TiOBr is illustrated by Figs. 3 and 4.  Figs. 3(a) and 3(b) provide Bose-corrected difference maps of S(Q,E) in the commensurate (T = 8K) and incommensurate (T = 37 K) spin-Peierls states.  While two branches of magnetic excitations can be observed in both of the low temperature phases, there are noticeable differences between the magnetic scattering at T = 8K and T = 37 K.  In particular, both the n = 1 and n = 2 triplet excitations appear weaker in the incommensurate spin-Peierls state, and both the first and second energy gaps appear to have partially filled in by T = 37 K.  This effect is also visible in the representative energy cuts provided in Figs. 3(c) and 3(d).  These cuts also help to demonstrate the magnetic origin of the inelastic features at $\sim$ 21 meV and 41 meV.  These two peaks are the only features which appear to decrease in intensity as the temperature increases, and thus can easily be distinguished from the phonon excitations which can be seen at $\Delta$E $\sim$ 15 meV, 25 meV, and 35 meV.

\begin{figure}
\includegraphics{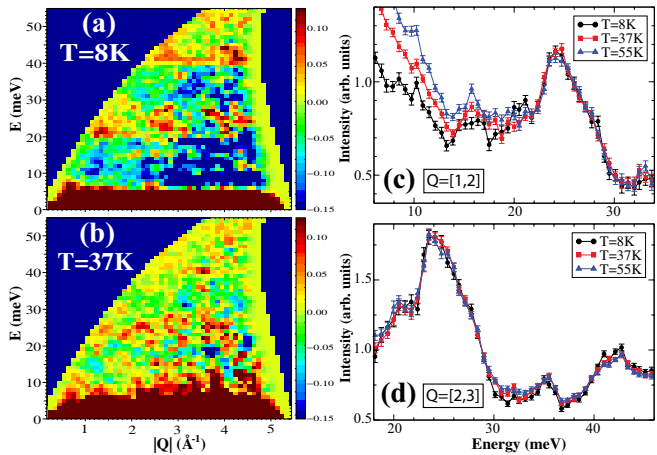}
\caption{(Color online) Temperature dependence of magnetic excitations in TiOBr.  Color contour maps of the inelastic scattering intensity, S(Q,E), are shown for (a) T = 8 K (within the commensurate spin-Peierls state) and (b) T = 37 K (within the incommensurate spin-Peierls state).  A Bose-corrected high temperature (T = 55 K) background has been subtracted from (a) and (b) in order to isolate the magnetic scattering.  Representative energy cuts, I(E), are shown for (c) {\bf Q} = [1, 2] {\AA}$^{-1}$ and (d) {\bf Q} = [2, 3] {\AA}$^{-1}$.  An empty can background has been subtracted from the data in (c) and (d) to eliminate scattering from the sample environment.}
\end{figure}

Fig. 4 illustrates the temperature dependence of the integrated intensity of the inelastic magnetic scattering.  Here we follow the temperature evolution of the inelastic scattering intensity at four different points: at the n = 1 and n = 2 triplet excitations (shown in Fig. 4(a)) and inside the first and second energy gaps (shown in Fig. 4(b)).  Integrated intensities were obtained by binning up all scattering intensity over ranges of:  {\bf Q} = 1 to 2 {\AA}$^{-1}$ and E = 20.5 to 21.8 meV (n = 1 triplet), {\bf Q} = 2 to 3 {\AA}$^{-1}$ and E = 40.4 to 42.1 meV (n = 2 triplet), {\bf Q} = 1 to 3 {\AA}$^{-1}$, E = 10 to 12 meV (first energy gap) and E = 30 to 32 meV (second energy gap).  The integrated intensities are all presented relative to the scattering at T = 8 K, at which point we assume that both the triplet excitations and the singlet-triplet energy gaps will be fully developed. Figs. 4(a) and 4(b) clearly show that with increasing temperature the n = 1 and n = 2 triplet excitations gradually lose intensity, while the energy gaps progressively gain intensity.  Thus, the magnetic excitations appear to weaken and broaden as the singlet-triplet gap slowly fills in with increasing temperature.

\begin{figure}
\includegraphics{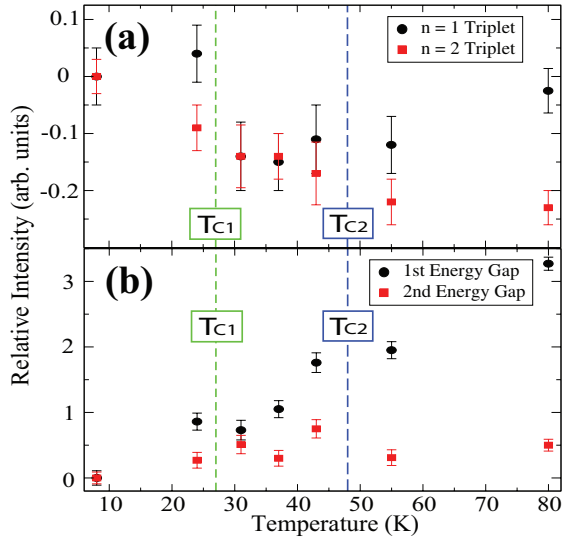}
\caption{(Color online) (a) Temperature dependence of the integrated scattering intensity for the n = 1 ($\Delta$E = [20.5, 21.8] meV, {\bf Q} = [1, 2] {\AA}$^{-1}$) and n = 2 ($\Delta$E = [40.4, 42.1] meV, {\bf Q} = [2, 3] {\AA}$^{-1}$) triplet excitations.  (b) Temperature dependence of the integrated scattering intensity within the first ($\Delta$E = [10, 12] meV, {\bf Q} = [1, 3] {\AA}$^{-1}$) and second ($\Delta$E = [30, 32] meV, {\bf Q} = [1, 3] {\AA}$^{-1}$) energy gaps.  All intensities are expressed relative to the scattering observed at T = 8 K.}
\end{figure}
   
Interestingly, the intensity of the n = 1 triplet excitation remains approximately constant and fully-developed throughout the commensurate spin-Peierls state.  It then drops rapidly near the discontinuous phase transition at T$_{C1}$ $\sim$ 27 K, and remains roughly constant through the incommensurate spin-Peierls state.  The intensity of the n = 2 triplet excitation gradually decreases through both the commensurate and incommensurate spin-Peierls phases, eventually reaching a plateau near the transition to the paramagnetic state at T$_{C2}$ $\sim$ 48 K.  

In conclusion, inelastic neutron scattering measurements reveal two branches of magnetic excitations in the commensurate (T $<$ 27 K) and incommensurate (27 K $<$ T $<$ 48 K) spin-Peierls phases of TiOBr, which can be understood as n = 1 and n = 2 triplet excitations from the singlet ground state.  The singlet-triplet energy gap was determined to be $E_{g}$ = 21.2 $\pm$ 1.0 meV, a result which is dramatically larger than the standard BCS prediction, but fully consistent with the anomalously large gap reported for TiOCl\cite{Imai, Baker, Lemmens_TiOCl, Caimi}.  The magnetic excitations in TiOBr exhibit relatively little dispersion, and are consistent with well-localized and weakly interacting excited triplets.  We hope these results will help to guide and inform future studies of these novel magnetic systems.

The authors would like to acknowledge S. H. Huang for sample preparation, and C. Stock, A. Aczel, and J.P.C. Ruff for helpful discussions.  This work was supported by NSERC of Canada and NSC of Taiwan under project No. NSC-98-2119-M-002-021.  Research at the SNS at ORNL was sponsored by the Scientific User Facilities Division, Office of Basic Energy Sciences, U.S. Dept. of Energy.  ORNL is managed by UT-Batelle, LLC, under contract DE-AC0500OR22725 for the U.S. Dept. of Energy.

%
%
%
%
%
%
%
%
%
%

\end{document}